\documentclass[epj]{svjour}
\usepackage{amsmath}
\usepackage{overpic}
\usepackage{graphics}
\usepackage{hyperref}
\usepackage{rotating}
\usepackage{cite}
\newcommand{\ket}[1]{|#1\rangle}
\newcommand{\bra}[1]{\langle #1|}

\begin{document}
\title{Pairwise Entanglement and Geometric Phase in High Dimensional Free-Fermion Lattice Systems}
\author{H. T. Cui \and Y. F. Zhang}
\institute{Department of Physics, Anyang Normal University, Anyang,
455000, China, \email{cuiht@aynu.edu.cn}}
\date{Received: \today / Revised version: \today}
\abstract{ The pairwise entanglement, measured by concurrence and
geometric phase in high dimensional free-fermion lattice systems
have been studied in this paper. When the system stays at the ground
state, their derivatives with the external parameter show the
singularity closed to the phase transition points, and can be used
to detect the phase transition in this model. Furthermore our
studies show for the free-fermion model that both concurrence and
geometric phase show the intimate connection with the correlation
functions. The possible connection between concurrence and geometric
phase has been also discussed. \PACS{
      {03.65.Vf} {Phases: geometric; dynamic or
topological;}
      {03.65.Ud} {Entanglement and quantum
nonlocality;} \and {05.70.Fh} {Phase transitions: general studies}
     }
}
\authorrunning{H.T. Cui, \textit{et.al.} }
\titlerunning{Pairwise Entanglement and Geometric Phase in Higher Dimensional Free-Fermion Lattice systems}
\maketitle

\section{introduction}
The understanding of quantum many-body effects based on the
fundamentals of quantum mechanics, has been raising greatly because
of the rapid development in quantum information theory\cite{afov07}.
Encouraged by the suggestion of Preskill\cite{preskill}, the
connection between the quantum entanglement and quantum phase
transition has been demonstrated first in 1D spin-$1/2$ $XY$
mode\cite{oo02}, and then was extended to more other spin-chain
systems and fermion systems (see Ref \cite{afov07} for a review).
Furthermore the decoherence of a simple quantum systems coupled with
the quantum critical environment has been shown the significant
features closed to the critical points \cite{ycw06, quan}. Regarding
these findings, the fidelity between the states across the
transition point has also been introduced to mark the happening of
the phase transitions \cite{zanardi}. These intricate connections
between quantum entanglement and phase transition in many-body
systems have sponsored great effort devoted to the understanding of
many-body effects from quantum information point\cite{afov07}. In
general quantum entanglement as a special correlation, is believed
to play an essential role for the many-body effects since it is well
accepted that the non-trivial correlation is at the root of
many-body effects. Although the ambiguity exists\cite{yang}, quantum
entanglement provides us a brand-new perspective into quantum
many-body effects. However the exact physical meaning of quantum
entanglement in many body systems remains unclear\cite{vedral07}.
Although the entanglement witnesses has been constructed in some
many-body systems\cite{wvb05}, a general and physical understanding
of quantum entanglement in many-body systems is still absent.

On the other hand, the geometric phase, which was first studied
systemically by Berry\cite{berry} and had been researched
extensively in the past 20 years\cite{gp}, recently has also been
shown the intimate connection to quantum phase
transitions\cite{cp05, zhu, hamma, cui06, plc06, cui08, hkh08}(or
see a recent review Ref.\cite{zhu08}). This general relation roots
at the topological property of the geometric phase, which depicts
the curvature of the Hilbert space, and especially has direct
relation to the property of the degeneracy in quantum systems. The
degeneracy in the many-body systems is critical in our understanding
of the quantum phase transition \cite{sachdev}. Thus the geometric
phase is another powerful tool for detecting the quantum phase
transitions. Moreover recently geometric phase has been utilized to
distinguish different topological phases in quantum Hall
systems\cite{shen}, in which the traditional phase transition theory
based on the symmetry-broken theory is not in
function\cite{Senthil}.

Hence it is very interesting to discuss the possible connection
between entanglement and geometric phase, since both issues show the
similar sensitivity to the happening of quantum phase transition.
Recently the connection between the entanglement entropy and
geometric phase has first been discussed with a special model in
strongly correlated systems; the geometric phase induced by the
twist operator imposed on the filled Fermi sphere, was shown to
present a lower bound for the entanglement entropy\cite{rh06}. This
interesting result implies the important relation between quantum
entanglement and geometric phase, and provides an possible
understanding of entanglement from the topological structure of the
systems. In another way the two-particle entanglement was also
important\cite{oo02}. Especially in spin-chain systems two-particle
entanglement is more popular and general because of the interaction
between spins, and furthermore the quantum information transferring
based on spin systems are generally dependent on the entanglement
between two particles\cite{ss05}. So it is a tempting issue to
extend this discussion to the universal two-particle entanglement
situation.

For this purpose the pairwise entanglement and geometric phase are
studied systemically in this paper.  Our discussion focuses on
nearest-neighbor entanglement in the ground state in free-Fermion
lattice systems because of the availability of the exact results. By
our own knowledge, this paper first presents the exact results of
entanglement and geometric phase in higher dimensional systems. In
Sec.\ref{model} the model will be provided, and the entanglement
measured by Wootter's concurrence is calculated by introducing
pseudospin operators. Furthermore the geometric phase is obtained by
imposing a globe rotation, and its relation with concurrence are
also discussed generally. In Sec.\ref{xy}, we discussed respectively
the concurrence and geometric phase in 2D and 3D cases. Finally, the
conclusion is presented in Sec.\ref{dc}.

\section{Model}\label{model}
The Hamiltonian for spinless fermions in lattice systems reads
\begin{equation}\label{h}
H=\sum_{\mathbf{ij}}^{L}c_{\mathbf{i}}^{\dagger}A_{\mathbf{ij}}c_{\mathbf{j}}+\frac{1}{2}(c_{\mathbf{i}}^\dagger
B_{\mathbf{ij}}c_{\mathbf{j}}^\dagger+ \text{h.c.}),
\end{equation}
in which $c_{\mathbf{i}}^{(\dagger)}$ is fermion
annihilation(creation) operator and $L$ is the total number of
lattice sites. The hermitity of $H$ imposes that matrix $A$ is
Hermit and $B$ is an anti-symmetry matrix. The configuration of
lattice does not matter for Eq. \eqref{h} since our discussion
focuses on the general case and available exact results. This model
obviously is solvable exactly and can be transformed into the free
Bogoliubov fermionic model. So it is also called free-fermion model.
By Jordan-Wigner transformation\cite{jw28} one can convert the
spin-chain systems into spinless fermions systems, in which the
physical properties can be readily determined. Therefore an
alternative approach is necessary by which one can treat solvable
fermion systems of arbitrary size. The model Eq. \eqref{h} serves
this purpose.

Without the loss of generality we assume  $A$ and $B$ to be
real\cite{lsm61}.  An important property of Eq. \eqref{h} is
\begin{equation}\label{s}
[H, \prod_{\mathbf{i}}^L(1-2c_{\mathbf{i}}^\dagger
c_{\mathbf{i}})]=0.
\end{equation}
This symmetry would greatly simplify the consequent calculation of
the reduced density matrix for two fermions. One can diagonalize Eq.
\eqref{h} by introducing linear transformation with real
$g_{\mathbf{ki}}$ and $h_{\mathbf{ki}}$\cite{lsm61}
\begin{equation}
\eta_{\mathbf{k}}=\frac{1}{\sqrt{L}}\sum_{\mathbf{i}}^{L}
g_{\mathbf{ki}}c_{\mathbf{i}}+h_{\mathbf{ki}}c_{\mathbf{i}}^\dagger,
\end{equation}
in which the normalization factor $1/\sqrt{L}$ have been included to
ensure the convergency under the thermodynamic limit. After some
algebra, the Hamiltonian Eq. \eqref{h} becomes
\begin{equation}
H=\sum_{\mathbf{k}} \Lambda_{\mathbf{k}}\eta_{\mathbf{k}}^\dagger
\eta_{\mathbf{k}} + \text{const}.
\end{equation}
in which $\Lambda_{\mathbf{k}}^2$ is the common eigenvalue of the
matrices $(A-B)(A+B)$ and $(A+B)(A-B)$ with the corresponding
eigenvectors $\phi_{\mathbf{ki}}=g_{\mathbf{ki}}+h_{\mathbf{ki}}$
and $\psi_{\mathbf{ki}}=g_{\mathbf{ki}}-h_{\mathbf{ki}}$
respectively (see Ref.\cite{lsm61} for details). The ground state is
defined as $\ket{g}$, which satisfies the relation
\begin{equation}
\eta_{\mathbf{k}}\ket{g}=0
\end{equation}

With respect to fermi operator $\eta_{\mathbf{k}}$, one has
relations
\begin{eqnarray}
\frac{1}{L}\sum_{\mathbf{i}} g_{\mathbf{ki}}g_{\mathbf{k'i}}+h_{\mathbf{ki}}h_{\mathbf{k'i}}&=&\delta^{(3)}_{\mathbf{k'k}}\nonumber\\
\frac{1}{L}\sum
_{\mathbf{i}}g_{\mathbf{ki}}h_{\mathbf{k'i}}+h_{\mathbf{ki}}g_{\mathbf{k'i}}&=&0
\end{eqnarray}
Furthermore the requirement that $\{\phi_k, \forall k\}$ and
$\{\psi_k, \forall k\}$ be normalized and complete, reinforce the
relations \cite{lsm61}
\begin{eqnarray}
\frac{1}{L}\sum_{\mathbf{k}}g_{\mathbf{ki}}g_{\mathbf{kj}}+h_{\mathbf{ki}}h_{\mathbf{kj}}&=&\delta_{\mathbf{ij}}\nonumber\\
\frac{1}{L}\sum_{\mathbf{k}}g_{\mathbf{ki}}h_{\mathbf{kj}}+h_{\mathbf{ki}}g_{\mathbf{kj}}&=&0
\end{eqnarray}
With the help of these formula above, one obtains
\begin{equation}
c_{\mathbf{i}}=\frac{1}{\sqrt{L}}\sum_{\mathbf{k}}g_{\mathbf{ki}}\eta_{\mathbf{k}}+h_{\mathbf{ki}}\eta_{\mathbf{k}}^\dagger,
\end{equation}
which would benefit our calculation for the correlation functions.

\subsection{Concurrence}
The concurrence, first introduced by Wootters\cite{wootters} for the
measure of two-qubit entanglement, is defined as
\begin{equation}
c=\max\{0, \lambda_1-\lambda_2-\lambda_3-\lambda_4\},
\end{equation}
in which $\lambda_i(i=1,2,3,4)$ are the square roots of eigenvalues
of matrix
$R=\rho(\sigma^y\otimes\sigma^y)\rho(\sigma^y\otimes\sigma^y)$ with
decreasing order. Then the critical step is to determine the
two-body reduced density operator $\rho$. The reduced density
operator $\rho_{\mathbf{ij}}$ for two spin-half particles labeled
$\mathbf{i, j}$ can be written generally as,
\begin{equation}
\rho_{\mathbf{ij}}=\text{tr}_{\mathbf{ij}}\rho=\frac{1}{4}\sum_{\alpha,
\beta=0}^{4}p_{\alpha,
\beta}\sigma^\alpha_{\mathbf{i}}\otimes\sigma^\beta_{\mathbf{j}},
\end{equation}
in which $\rho$ is the density matrix for the whole system and
$\sigma^0$ is the $2\times2$ unity matrix and
$\sigma^{\alpha}(\alpha=1,2,3)$ are the Pauli operators $\sigma^x,
\sigma^y, \sigma^z$, which also the generators of $SU(2)$ group.
$p_{\alpha\beta}=\text{tr}[\sigma^{\alpha}_{\mathbf{i}}\sigma^\beta_{\mathbf{j}}\rho_{\mathbf{ij}}]
=\langle\sigma^{\alpha}_{\mathbf{j}}\sigma^\beta_{\mathbf{j}}\rangle$
is the correlation function. With the symmetry Eq. \eqref{s}, one
can verify that only $p_{00}, p_{03}, p_{30}, p_{11}, p_{22},
p_{33}, p_{12}, p_{21}$ are not vanishing. After some efforts, one
obtain
\begin{equation}
c=\max\{0, c_I, c_{II}\},
\end{equation}
in which
\begin{eqnarray}\label{c}
c_I&=&\frac{1}{2}[\sqrt{(p_{11}+p_{22})^2+(p_{12}-p_{21})^2}\nonumber\\
&&-\sqrt{(1+p_{33})^2-(p_{30}+p_{03})^2}]\nonumber\\
c_{II}&=&\frac{1}{2}[|p_{11}-p_{22}|-\sqrt{(1-p_{33})^2-(p_{30}-p_{03})^2}].
\end{eqnarray}

In order to obtain the reduced density operator for two fermions, it
is crucial to construct $SU(2)$ algebra for the fermions in lattice
systems. In 1D case, the Jordan-Wigner (JW) transformation is
available\cite{jw28, cp05, zhu, lsm61}. For higher dimension cases
the JW-like transformation has been constructed by different
methods\cite{jw}. However the transformation is very complex and the
calculation is difficult. Hence instead of a general calculation, we
focus on the nearest neighbor two lattices in this paper. In this
situation, the $SU(2)$ algebra can be readily constructed
\begin{eqnarray}\label{jw2}
\sigma_{\mathbf{i}}^+=(\sigma_{\mathbf{i}}^x+i\sigma_{\mathbf{i}}^y)/2=c^{\dagger}_{\mathbf{i}}\nonumber\\
\sigma_{\mathbf{i}}^-=(\sigma_{\mathbf{i}}^x-i\sigma_{\mathbf{i}}^y)/2=c_{\mathbf{i}}\nonumber\\
\sigma_{\mathbf{i}}^z=2c_{\mathbf{i}}^{\dagger} c_{\mathbf{i}}-1\nonumber\\
\sigma_{\mathbf{i}+1}^+=(\sigma_{\mathbf{i}+1}^x+i\sigma_{\mathbf{i}+1}^y)/2=(2c_{\mathbf{i}}^{\dagger}c_{\mathbf{i}}-1)c^{\dagger}_{\mathbf{i}+1}\nonumber\\
\sigma_{\mathbf{i}+1}^-=(\sigma_{\mathbf{i}+1}^x-i\sigma_{\mathbf{i}+1}^y)/2=(2c_{\mathbf{i}}^{\dagger}c_{\mathbf{i}}-1)c_{\mathbf{i}+1}\nonumber\\
\sigma_{\mathbf{i}+1}^z=2c_{\mathbf{i}+1}^{\dagger}c_{\mathbf{i}+1}-1
\end{eqnarray}
in which $\mathbf{i}+1$ denotes the nearest neighbor lattice for
site $\mathbf{i}$. This point can be explained as the following. The
difficulty for the JW transformation in higher dimension case comes
from the absence of a natural ordering of particles. However when
one focuses on the nearest neighbored particle, this difficulty does
not appear since for a definite direction the nearest neighbor
particle is unique (for non-nearest neighbored case one have to
consider the effect from the other particles). Then the correlation
functions for the ground state are in this case
\begin{eqnarray}\label{cf}
p_{00}&=&1, p_{30}=1-\frac{2}{L}\sum_{\mathbf{k}}h_{\mathbf{ki}}^2; p_{03}=1-\frac{2}{L}\sum_{\mathbf{k}}h_{\mathbf{k(i+1)}}^2;\nonumber\\
p_{11}&=&\frac{2}{L}\sum_{\mathbf{k}}(h_{\mathbf{ki}}-g_{\mathbf{ki}})(h_{\mathbf{k(i+1)}}+g_{\mathbf{k(i+1)}});\nonumber\\
p_{22}&=&\frac{2}{L}\sum_{\mathbf{k}}(h_{\mathbf{ki}}+g_{\mathbf{ki}})(h_{\mathbf{k(i+1)}}-g_{\mathbf{k(i+1)}})\nonumber\\
p_{33}&=&(1-\frac{2}{L}\sum_{\mathbf{k}}h^2_{\mathbf{ki}})(1-\frac{2}{L}\sum_{\mathbf{k}}h^2_{\mathbf{k(i+1)}})\nonumber\\
&&+\frac{4}{L^2} \sum_{\mathbf{k,
k'}}h_{\mathbf{ki}}h_{\mathbf{k(i+1)}}g_{\mathbf{k'i}}g_{\mathbf{k'(i+1)}}-h_{\mathbf{ki}}g_{\mathbf{ki}}h_{\mathbf{k'(i+1)}}
g_{\mathbf{k'(i+1)}}\nonumber\\
p_{12}&=&p_{21}=0
\end{eqnarray}

\subsection{Geometric Phase}
Following the method in Refs.\cite{cp05, zhu}, one can introduce a
globe rotation $R(\phi)=\exp[i\phi\sum_i c_{\mathbf{i}}^\dagger
c_{\mathbf{i}}]$ to obtain the geometric phase(GP). Then we have
Hamiltonian with parameter $\phi$
\begin{equation}\label{hphi}
H(\phi)=\sum_{\mathbf{ij}}^{L}c_{\mathbf{i}}^{\dagger}A_{ij}c_{\mathbf{j}}+\frac{1}{2}(c_{\mathbf{i}}^\dagger
B_{\mathbf{ij}}c_{\mathbf{j}}^\dagger e^{2i\phi}+\text{h.c.}),
\end{equation}
and the ground state becomes $\ket{g(\phi)}=R(\phi)\ket{g}$. GP is
defined as \cite{berry}
\begin{eqnarray}\label{gp}
\gamma_g&=&-i\int
d\phi\bra{g(\phi)}\frac{\partial}{\partial\phi}\ket{(\phi)}\nonumber\\&=&\frac{\phi}{L}\sum_{\mathbf{i}}\sum_{\mathbf{k}}h_{\mathbf{ki}}^2
\end{eqnarray}
Regarding to Eq.\eqref{hphi}, one only require $\phi=\pi$ for a
cycle evolution. Hence one has
$\gamma_g=\frac{\pi}{L}\sum_i\sum_{\mathbf{k}}h_{\mathbf{ki}}^2=\frac{1}{L}\sum_{\mathbf{i}}\gamma_{g\mathbf{i}}$.

\subsection{GP vs. Concurrence}
At a glance of Eq.\eqref{c} and Eq.\eqref{gp}, GP and concurrence
both are related directly to correlation functions. Hence it is
tempting to find the relation between the two quantities, which
would benefit to the understanding of the physical meaning of
concurrence.

According to Eqs.\eqref{c} and \eqref{cf}, the following inequality
can be obtained (see Appendix for details of calculations)
\begin{eqnarray}\label{cineq}
c_{I}&\le&\frac{1}{L\pi}(\gamma_{g\mathbf{i}}+\gamma_{g(\mathbf{i+1})})-\sqrt{(1+p_{33})^2-(p_{30}+p_{03})^2}
\nonumber\\
c_{II}&\le&1+\frac{1}{L\pi}(\gamma_{g\mathbf{i}}-\gamma_{g\mathbf{(i+1)}})-
\frac{1}{2L^2\pi^2}(\gamma_{g\mathbf{i}}-\gamma_{g\mathbf{(i+1)}})^2
\end{eqnarray}
For the first inequality, a much tighter bound is difficult to find.
While if the average of $c_{II}$ over all site $\mathbf{i}$ is
considered, $c_{II}\le1-
\frac{1}{2L^3\pi^2}\sum_i(\gamma_{g\mathbf{i}}-\gamma_{g\mathbf{(i+1)}})^2$.
Fortunately in the following examples $c_I$ is always negative.
Although the existence of this defect, in our own points, the
relation between GP and concurrence have been displayed genuinely
from the inequality above.

\section{GP and Concurrence in Higher Dimensional $XY$ model}\label{xy}
The previous section presents the general discussion of GP and
concurrence in free fermion lattice system Eq.\eqref{h}. In this
section a concrete model would be checked explicitly, of which the
Hamiltonian is
\begin{equation}\label{hd}
H=\sum_{\langle \mathbf{i,j} \rangle}[c_{\mathbf{i}}^\dagger
c_{\mathbf{j}}-\gamma(c_{\mathbf{i}}^\dagger
c_{\mathbf{j}}^\dagger+\text{h.c.})]-2\lambda\sum_{\mathbf{i}}
c_{\mathbf{i}}^\dagger c_{\mathbf{i}},
\end{equation}
in which $\langle \mathbf{i,j} \rangle$ denotes the nearest-neighbor
lattice sites and $c_{\mathbf{i}}$ is fermion operator. This
Hamiltonian, first introduced in Ref.\cite{li06}, depicts the
hopping and pairing between nearest-neighbor sites in hypercubic
lattice systems, in which $\lambda$ is the chemical potential and
$\gamma$ is the pairing potential. Eq.\eqref{hd} could be considered
as a $d$-dimensional generalization of 1D XY model. However for
$d>1$ case, this model shows different phase features \cite{li06}.

The Hamiltonian can be diagonalized by introducing the
$d$-dimensional Fourier transformation with periodic boundary
condition in momentum space \cite{li06}
\begin{equation}
H=\sum_{\mathbf{k}}2t_{\mathbf{k}} c_{\mathbf{k}}^\dagger
c_{\mathbf{k}}-i\Delta_{\mathbf{k}}(c_{\mathbf{k}}^\dagger
c_{-\mathbf{k}}^\dagger - \text{h.c.}),
\end{equation}
in which $t_{\mathbf{k}}=\sum_{\alpha=1}^d\cos k_\alpha-\lambda$ and
$\Delta_{\mathbf{k}}=\gamma\sum_{\alpha=1}^d\sin k_{\alpha}$. With
the help of Bogoliubov transformation, one obtains
\begin{equation}
H=\sum_{\mathbf{k}}2\Lambda_{\mathbf{k}}\eta_{\mathbf{k}}^\dagger\eta_{\mathbf{k}}+\text{const}.
\end{equation}
in which
$\Lambda_{\mathbf{k}}=\sqrt{t_{\mathbf{k}}^2+\Delta_{\mathbf{k}}^2}$.
Based on the degeneracy of the eigenenergy $\Lambda_{\mathbf{k}}=0$,
the phase diagram can be determined clearly\cite{li06}; When $d=2$,
the phases diagram should be identified as two different situations;
for $\gamma=0$, the degeneracy of the ground state occurs when
$\lambda\in[0, 2]$, whereas the gap above the ground state is
non-vanishing for $\lambda>2$. However for $\gamma\neq0$ three
different phases can be identified as $\lambda=0$, $\lambda\in(0,
2]$ and $\lambda>2$. The first two phases correspond to case that
the energy gap above the ground state vanishes, whereas not for
$\lambda>2$. One should note that $\lambda=0$ means a well-defined
Fermi surface with $k_x=k_y\pm\pi$, whose symmetry is lowered by the
presence of $\lambda$ term. For $d=3$ two phases can be identified
as $\lambda\in[0,3]$ with the vanishing energy gap above the ground
state and $\lambda>3$ with a non-vanishing energy gap above ground
state. In a word the critical points can be identified as
$\lambda_c=d (d=1,2,3)$ for any anisotropy of $\gamma$, and
$\lambda=0$ for $d=2$ with $\gamma\neq0$. One should note that since
the $\gamma^2$ dependence of $\Lambda_{\mathbf{k}}$, the sign of
$\gamma$ does not matter. Hence the plots below are only for
positive $\gamma$.

The correlation functions between nearest-neighbor lattice sites
would play a dominant role in the transition between different
phases because of the nearest-neighbor interaction, similar to the
case in XY model \cite{oo02}. Then it is expected  that the pairwise
entanglement is significant in this model. In the following,
concurrence for the nearest-neighbor sites of ground state is
calculated for $d=2, 3$ respectively. The geometric phase of ground
state is also calculated by imposing a globe rotation $R(\phi)$. our
calculation shows that both quantities show interesting singularity
closed to the boundary of different phases.

\subsection{Concurrence}
For $d>1$ case, the nearest-neighbor lattice sites appear in
different directions. In order to eliminate the dependence of
orientations, the calculation of correlation functions
Eqs.\eqref{cf} is implemented by averaging in all directions. With
the transformation Eq.\eqref{jw2}, one can determine under the
thermodynamic limit
\begin{eqnarray}
p_{11}&=&\frac{1}{d(2\pi)^d}\int_{-\pi}^{\pi}\prod_{\alpha}^d
dk_\alpha(\Delta_k\sum_{\alpha=1}^d\sin
k_{\alpha}-t_k\sum_{\alpha=1}^d\cos k_\alpha )/\Lambda_k\nonumber\\
p_{22}&=&-\frac{1}{d(2\pi)^d}\int_{-\pi}^{\pi}\prod_{\alpha}^d
dk_\alpha(\Delta_k\sum_{\alpha=1}^d\sin
k_{\alpha}+t_k\sum_{\alpha=1}^d\cos k_\alpha )/\Lambda_k\nonumber\\
p_{12}&=&p_{21}=0\nonumber\\
p_{03}&=&p_{30}=p_3=\frac{1}{(2\pi)^d}\int_{-\pi}^{\pi}\prod_{\alpha}^d
dk_\alpha \frac{t_k}{\Lambda_k}\nonumber\\
p_{33}&=&p_3^2-(\frac{p_{11}+p_{22}}{2})^2+(\frac{p_{11}-p_{22}}{2})^2
\end{eqnarray}

\begin{figure}
\center
\begin{overpic}[width=6cm]{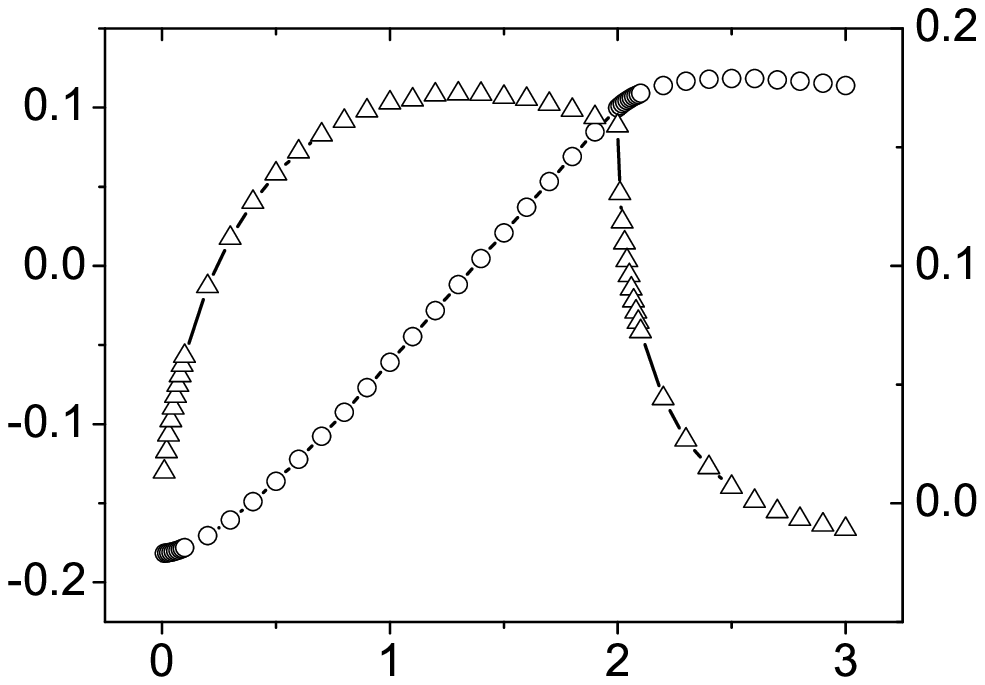}
\put(-5, 40){\large$c_{II}$} \put(50 , 0){\large$\lambda$}\put(97,
50){\large \begin{rotate}{-90}$\partial
c_{II}/\partial\lambda$\end{rotate}}
\end{overpic}\vspace{2em}
\caption{\label{2c} $c_{II}$ ($\bigcirc$) and its derivative with
$\lambda$ ($\triangle$) vs. $\lambda$ when $d=2$. We have chosen
$\gamma=1$ for this plot.}
\end{figure}

$d=2$ Our calculation shows that $c_{I}$ is negative. So in Fig.
\ref{2c}, only $c_{II}$ and its derivative with  $\lambda$ are
numerically  illustrated. In order to avoid the ambiguity because of
the cutoff in the definition of concurrence, the derivative of
$c_{II}$ with $\lambda$ is depicted in all region whether $c_{II}$
positive or not\cite{yang}. Obviously the singularity for $\partial
c_{II}/\partial\lambda$ can be found at the point $\lambda=0, 2$
respectively, which are consistent with our knowledge about phase
transitions.

\begin{figure}[tb]\center
\begin{overpic}[width=6cm]{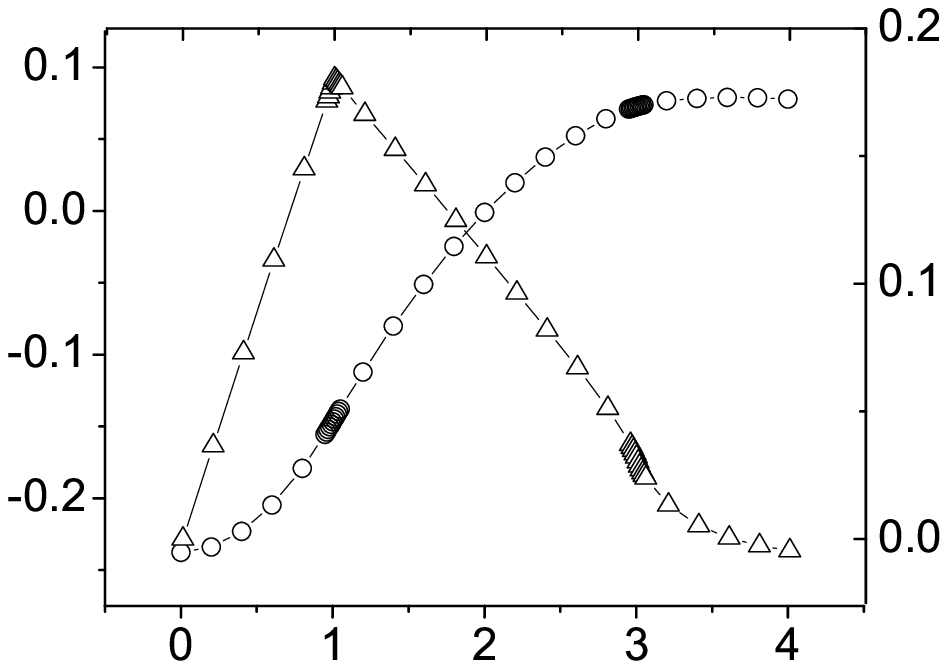}
\put(15, 60){(a)}\put(-5, 40){\large $c_{II}$} \put(45, 0){\large
$\lambda$} \put(94, 50){\large \begin{rotate}{-90}$\partial
c_{II}/\partial\lambda$\end{rotate}}
\end{overpic}\vspace{-2em}
\\
\begin{overpic}[width=4.5cm]{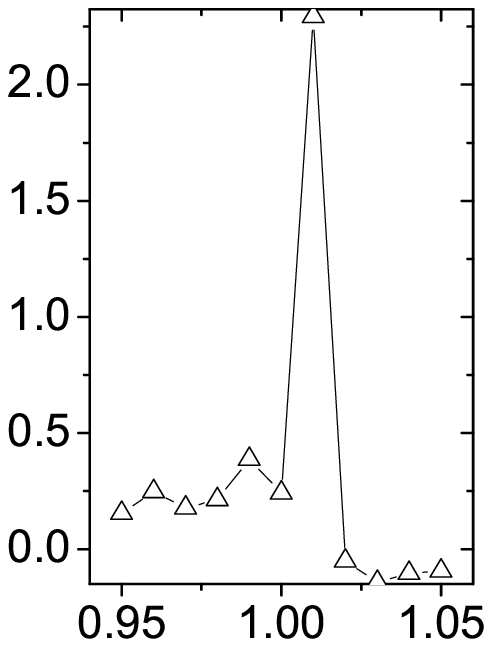}
\put(18,78){(b)} \put(5, 35){\large \begin{rotate}{90}$\partial^2
c_{II}/\partial\lambda^2$\end{rotate}} \put(40, 0){\large $\lambda$}
\end{overpic}\hspace{-4em}
\begin{overpic}[width=4.8cm]{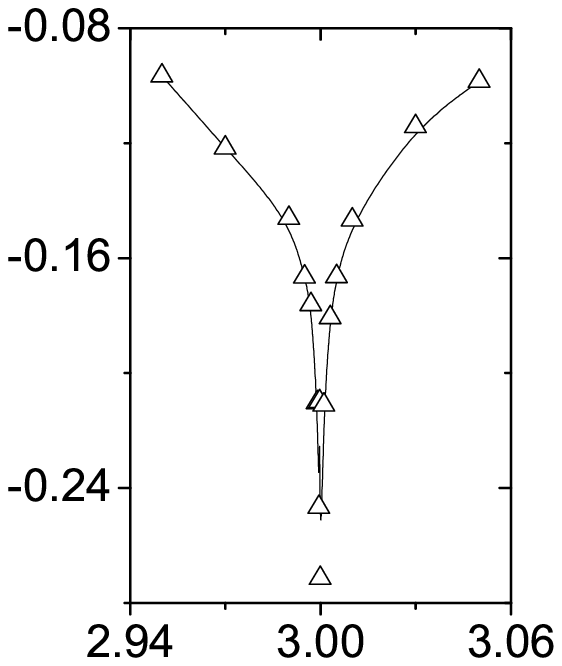}
\put(22, 78){(c)}\put(40, 0){\large $\lambda$}
\end{overpic}
\caption{\label{3c} $c_{II}$ ($\bigcirc$) and its derivative with
$\lambda$ ($\triangle$) (a) vs. $\lambda$ when $d=3$. We have chosen
$\gamma=1$ for this plot. The second derivative of $c_{II}$ with
$\lambda$ are also displayed in this plot and focus the points
closed to $\lambda=1$ (b) and $\lambda=3$ (c).}
\end{figure}

$d=3$  Similar to the case of $d=2$, our calculation shows $c_I<0$.
Only $c_{II}$ and its derivative with $\lambda$ are numerically
displayed in Fig.\ref{3c}. Different from the case of $d=2$, no
singularity of the first derivative of $c_{II}$ with $\lambda$ is
found at $\lambda=3$. While a cusp appears at $\lambda=1$. A further
calculation demonstrates that the second derivative of $c_{II}$ is
divergent genuinely at exact $\lambda=3$, as shown in
Figs.\ref{3c}(c). which means the phase transition at this points.
Furthermore our numerical calculations show that $\partial^2
c_{II}/\partial\lambda^2$ is finite at $\lambda=1$, as shown in
Figs.\ref{3c}(b). Hence one cannot attribute this feature to the
phase transition. The similar feature has been found in the previous
studies \cite{oo02, yang, gu}. However the underlying physical
reason is unclear in general. But this special feature is not unique
for concurrence; van Hove singularity in solid state physics
displays the similar feature, which is because of the vanishing of
the moment-gradient of the energy. Although we cannot established
the direct relation between these two issues because of the bad
definition of the moment-gradient of the energy when degeneracy
happening, we affirm that this feature is not an accident and the
underlying physical reason is still to be found.

In a word the discussion above first demonstrates the exact
connection between concurrence and quantum phase transitions in
high-dimensional many body systems. However a question is still
open; what the physical interpretation of concurrence is in
many-body systems. In this study, we includes the negative part of
$c_{II}$ to identify the phase diagram in free-fermion systems. In
general, it is believed that the negative $c_{II}$ means no
entanglement between two particles and then include no any useful
information about state. But from the discussion one can note that
the omission of the negative part of $c_{II}$ would lead to
incorrect results. Moreover, for $\gamma=0$, our calculations show
that $c_I, c_{II}$ always are zero, and so one cannot obtain any the
phase transition information from pair wise entanglement in this
case. Further discussions will be presented in the final part of
this paper.

\subsection{Geometric Phase} Geometric phase manifests the
structure of Hilbert in the system and has intimate relation to the
degeneracy. GP, defined in Eq. \eqref{gp} by imposing a globe
rotation $R(\phi)$ on ground state $\ket{g}$ is calculated in this
section. After some algebra, one obtains
\begin{equation}
\gamma_g=\frac{\pi}{2(2\pi)^d}\int_{-\pi}^{\pi}\prod_{\alpha=1}^ddk_\alpha
(1-\frac{t_k}{\Lambda_k}).
\end{equation}

$d=2$  In Fig.\ref{2g}, $\gamma_g$ and its derivative with $\lambda$
are displayed explicitly. Obviously one notes that
$\partial\gamma_g/\partial\lambda$ shows the singularity closed to
$\lambda=0, 2$, which are exactly the phase transition points of
Hamiltonian Eq.\eqref{hd}. An interesting observation is that closed
to these points, both GP and concurrence $c_{II}$ show the similar
behaviors.

\begin{figure}[tb]\center
\begin{overpic}[width=6cm]{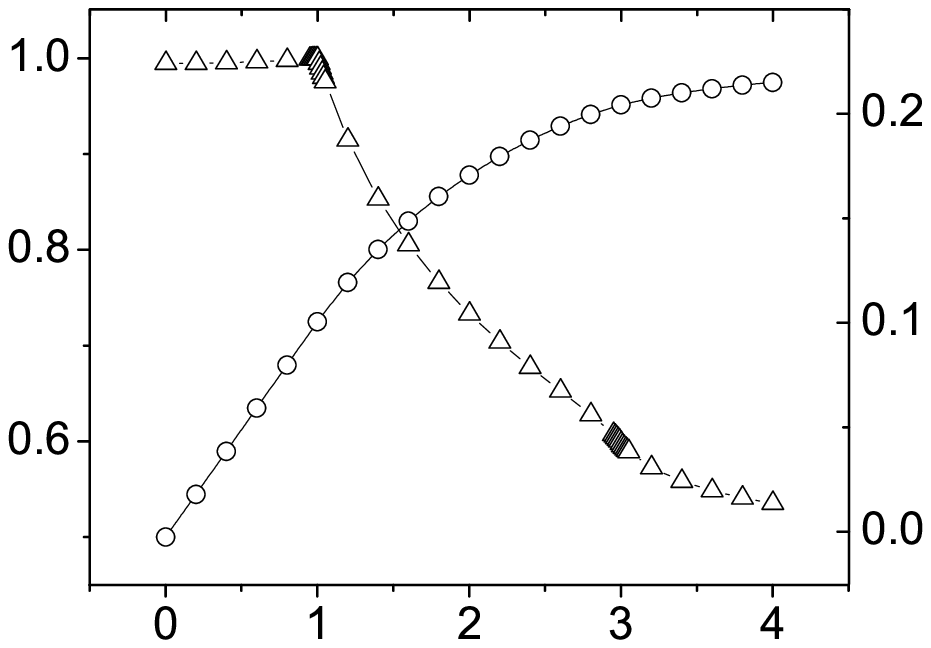}
\put(13, 60){(a)}\put(0, 35){\large
\begin{rotate}{90}$\gamma_g/\pi$\end{rotate}} \put(50, 0){\large $\lambda$}
\put(93, 52){\large
\begin{rotate}{-90}$\partial
\gamma_g/\pi\partial\lambda$\end{rotate}}
\end{overpic}\vspace{-2em}
\begin{overpic}[width=4cm]{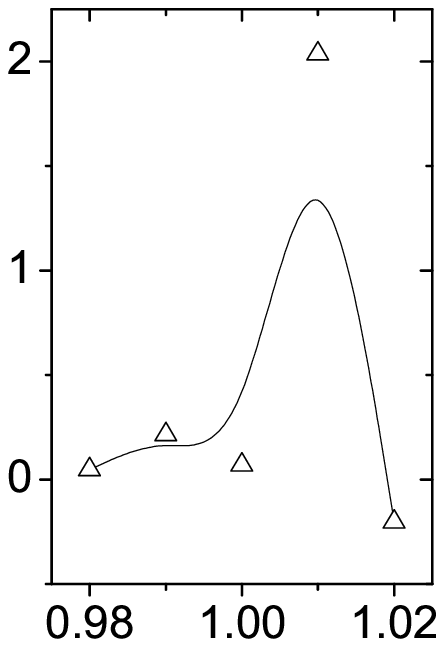}
\put(13,77){(b)} \put(5, 35){\large \begin{rotate}{90}$\partial^2
\gamma_g/\pi\partial\lambda^2$\end{rotate}} \put(35, 0){\large
$\lambda$}
\end{overpic}\hspace{-3em}
\begin{overpic}[width=4.6cm]{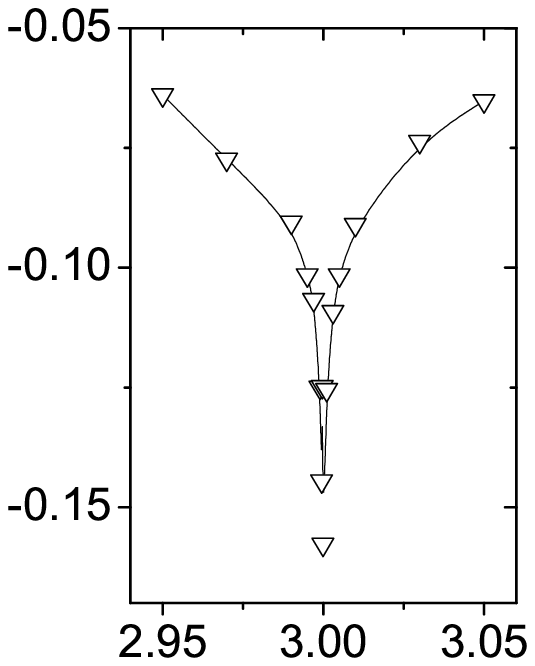}
\put(22, 78){(c)}\put(42, 0){\large $\lambda$}
\end{overpic}
\caption{\label{3g}$\gamma_g$ ($\bigcirc$) and its derivative with
$\lambda$ ($\triangle$) (a) vs. $\lambda$ when $d=3$. We have chosen
$\gamma=1$ for this plot. The second derivative` of $\gamma_g$ with
$\lambda$ are also displayed in this plot and focus the points
closed to $\lambda=1$ (b) and $\lambda=3$ (c).}
\end{figure}

$d=3$  GP and its derivative are plotted explicitly in
Fig.\eqref{3g}. One should note that there is a platform below
$\lambda=1$ for $\partial \gamma_g/\partial\lambda$, as shown in
Fig.\ref{3g}(a), but a further calculation shows that
$\partial^2\gamma_g/\partial\lambda^2$ is continued
(Fig.\ref{3g}(b)) and $\partial\gamma_g/\partial\lambda$ has no
divergency at this point. This phenomena is very similar to the case
of concurrence (see Fig.\ref{3c}(b, c)).  As expected,
$\partial^2\gamma_g/\partial\lambda^2$ is divergent at exact
$\lambda=3$, which means a phase transition happens  at this point.
Together with respect of the case of $d=2$, it makes us a suspect
that GP and concurrence in our model have the same physical
origination.

Furthermore for $\gamma=0$, GP fails to mark the phase transition
too. This is similar to the case of concurrence, but has different
physical reason. The further discussion is presented in the next
section.

\section{Discussion and Conclusions}\label{dc}
\begin{figure}\center
\begin{overpic}[width=4.4cm]{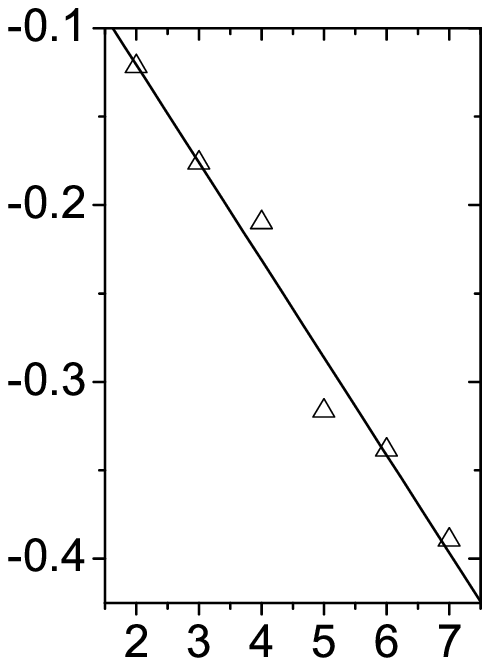}
\put(55,78){(a)} \put(5, 35){\large \begin{rotate}{90}$\partial^2
c_{II}\partial\lambda^2$\end{rotate}} \put(50, 0){\large $\log
|\lambda-\lambda_c|/ \lambda_c$}
\end{overpic}\hspace{-2em}
\begin{overpic}[width=4.6cm]{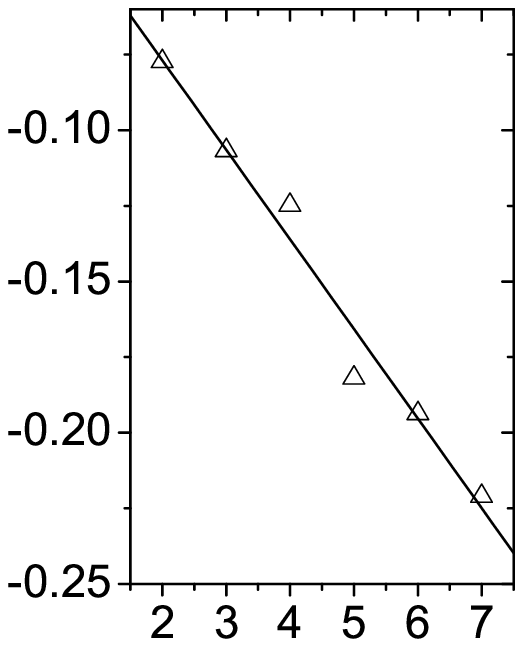}
\put(57, 78){(b)}\put(5, 35){\large \begin{rotate}{90}$\partial^2
\gamma_g/\pi\partial\lambda^2$\end{rotate}}
\end{overpic}
\caption{\label{5}The scaling of GP and concurrence for 3D case
closed to the critical point $\lambda_c=3$. We have chosen
$\gamma=1$ for this plot.}
\end{figure}

The pairwise entanglement and geometric phase for ground state in
$d$-dimensional ($d=2,3$) free-fermion lattice systems are discussed
in this paper. By imposing the transformation Eq.\eqref{jw2}, the
reduce two-body density matrix for the nearest neighbor particles
can be determined exactly for any dimension, and the concurrence is
also calculated explicitly. Furthermore geometric phase for ground
state, obtained by introducing a globe rotation $R(\phi)$, has also
been calculated. Given the known results for XY model \cite{oo02,
cp05, zhu}, our calculations show again that both GP and concurrence
display intimate connection with the phase transitions. Moreover an
inequality relation between concurrence and geometric phase is also
presented in Eq. \eqref{cineq}. The similar scaling behaviors at the
transition point $\lambda=3$ has also been shown in Figs. \ref{5}.
These facts strongly mean the intimate connection between the two
items. This point can be understand by noting that both of them are
connected to the correlation functions, as shown in Eqs. \eqref{c}
and \eqref{gp}.

An interesting point in our study is that in order to obtain all
information of phase diagram in model Eq.\eqref{hd}, the negative
part of $c_{II}$ has to be included to avoiding the confusion
because of the mathematical cutoff in the definition of
concurrence\cite{yang}. In general, it is well accepted that the
negative part of $c_{II}$ gives no any information of quantum
pairwise entanglement, and then is considered to be meaningless.
However, in our calculation, the negative part of $c_{II}$ appears
as an indispensable consideration to obtain the correct phase
information. This point means that the pairwise entanglement does
not provide the all information about the system since the two-body
reduced density operator throw away  much information.

\begin{figure}[tpb]\center
\begin{overpic}[width=6cm]{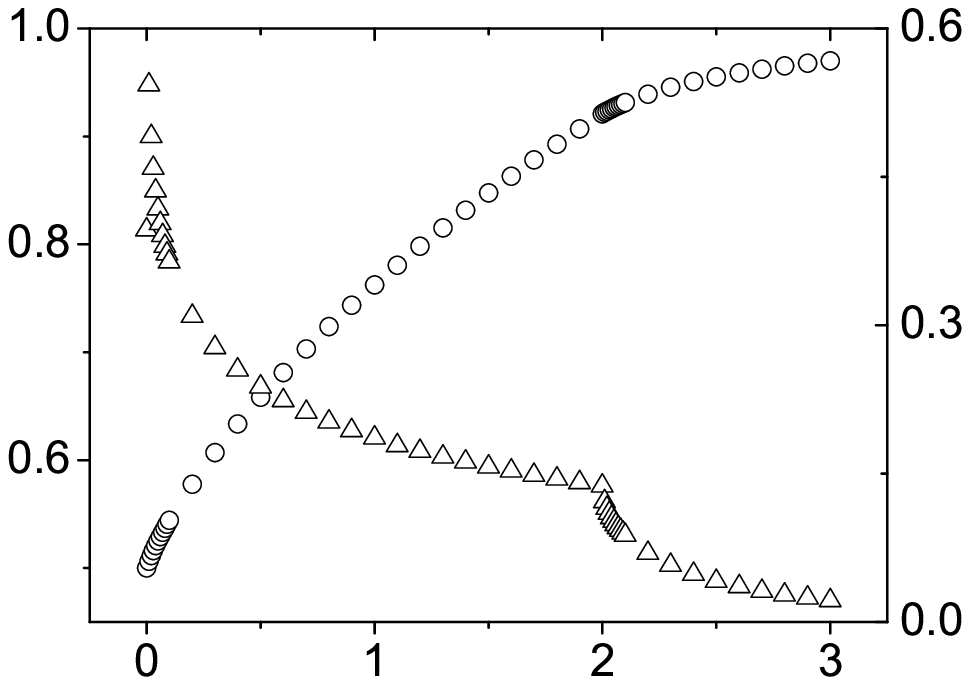}
\put(0, 35){\large  \begin{rotate}{90}$\gamma_g/\pi$\end{rotate}}
\put(50 , 0){\large $\lambda$}\put(98, 53){\large
\begin{rotate}{-90}$\partial
\gamma_g/\pi\partial\lambda$\end{rotate}}
\end{overpic}
\caption{\label{2g}$\gamma_g$ ($\bigcirc$) and its derivative with
$\lambda$ ($\triangle$) vs. $\lambda$ when $d=2$. We have chosen
$\gamma=1$ for this plot.}
\end{figure}

As for the geometric phase, defined in Eq. \eqref{gp}, it is obvious
that $\gamma_g$ can tell us the happening of phase transition at the
point, where $\gamma_g$ display some kinds of singularity. However
it cannot distinguished the degenerate region from the
nondegenerate, as shown in Figs. \ref{2g} and \ref{3g}. Recently GP
imposing by the twist operator in many-body systems is introduced as
an order parameter to distinguish the phases \cite{cui08, hkh08}.
For the free-fermion lattice system, this GP have also calculated
and shows the intimate connection with the vanishing of energy gap
above the ground state. However the boundary between the two
different phases becomes obscure with the increment of
dimensionality in that discussion \cite{cui08}, and moreover it
cannot distinguish the phase transition not come from the degeneracy
of the ground-state energy. While the geometric phase imposing by
the globe rotation $R(\phi)$ clearly demonstrate the existence of
this kind of phase transition, as shown in Fig.\ref{2g}, whether
originated from the degeneracy or not. In fact this point can be
understood by noting the intimate relation between $\gamma_g$ and
correlation functions. It maybe hint that one has to find different
methods for different many-body systems to identify the phase
diagram.

Although the intimate relationship of concurrence and GP with phase
transitions in the model Eq.\eqref{hd}, a exceptional happens when
$\gamma=0$, in which $c_I, c_{II}$ are zero and GP is a constant
independent of $\lambda$.  From Eq.\eqref{hd}, $\gamma=0$ means the
hopping of particles is dominant, and the position of particle
becomes meaningless. Since the calculation of concurrence depend on
the relative position of lattice site,  the pairwise entanglement is
disappearing. However one could introduce the spatial entanglement
to detect the phase transition in this case\cite{hav07}. For GP,
$\gamma=0$ means the emergency of new symmetry. One can find
$[\sum_{\mathbf{i}}c^\dagger_{\mathbf{i}}c_{\mathbf{i}}, H]=0$ in
this case, which leads to the failure of $R(\phi)$ for construction
of nontrivial GP.

Finally we try to transfer two viewpoints in this paper. One is that
concurrence and geometric phase can be used to mark the phase
transition in many-body systems since both of them are intimately
connected to the correlation functions. The other is that
concurrence and the geometric phase are connected directly by the
inequality Eq. \eqref{cineq}. Then it is interesting to extend this
relation to multipartite entanglement in the future works, which
would be helpful to establish the physical understanding of
entanglement.

\begin{acknowledgement}
The author (Cui) would appreciate the help from Dr. Kai Niu (DLUT)
and Dr. Chengwu Zhang (NJU) in the numerical calculations and
permission of the usage of their powerful computers. We also thank
greatly the enlightening discussion with Dr. Chong Li (DLUT).
Especially we thank the first referee for his/her important hint for
the van Hove singularity. This work is supported by the Special
Foundation of Theoretical Physics of NSF in China, Grant No.
10747159.
\end{acknowledgement}

\section*{APPENDIX}
For the first inequality, one should note
\begin{eqnarray}
&&|p_{11}+p_{22}|\nonumber\\&=&\frac{4}{L}|\sum_{\mathbf{k}}h_{\mathbf{ki}}h_\mathbf{k(i+1)}|\le\frac{4}{L}
\sum_{\mathbf{k}}|h_{\mathbf{ki}}h_\mathbf{k(i+1)}|
\nonumber\\&\le&\frac{2}{L}\sum_{\mathbf{k}}(h^2_{\mathbf{ki}}+h^2_\mathbf{k(i+1)})=\frac{2}{L\pi}(\gamma_{g\mathbf{i}}+\gamma_{g(\mathbf{i+1})}).
\end{eqnarray}
From inequality $\sqrt{x^2-y^2}\ge|x|-|y| (|x|>|y|)$, one reduces
\begin{equation}
\sqrt{(1+p_{33})^2-(p_{30}+p_{03})^2}]\ge|1+p_{33}|-|p_{30}+p_{03}|.
\end{equation}
Then one obtains
\begin{equation}
c_{I}\le\frac{1}{L\pi}(\gamma_{g\mathbf{i}}+\gamma_{g(\mathbf{i+1})})+\frac{1}{2}(|p_{30}+p_{03}|-|1+p_{33}|).
\end{equation}
However a much tighter bound is difficult to decide because of the
complexity of $p_{33}$.

For the second inequality, it can be obtained easily by observing
\begin{eqnarray}
p_{33}\le1-\frac{1}{L^2\pi^2}(\gamma_{g\mathbf{i}}^2-\gamma_{g(\mathbf{i+1})}^2),
\end{eqnarray}
in which we have used the relation $2ab\le a^2+b^2$. Then $1-p_{33}$
is non-negative and
\begin{eqnarray}
c_{II}&=&\frac{2}{L}\sum_{\mathbf{k}}|h_{\mathbf{ki}}g_\mathbf{k(i+1)}|+\frac{p_{33}-1}{2}\nonumber\\
&\le&\frac{1}{L}\sum_{\mathbf{k}}(h^2_{\mathbf{ki}}+g^2_\mathbf{k(i+1)})-
\frac{1}{2L^2\pi^2}(\gamma_{g\mathbf{i}}-\gamma_{g(\mathbf{i+1})})^2\nonumber\\
&\le&1+\frac{1}{L\pi}(\gamma_{g\mathbf{i}}-\gamma_{g\mathbf{(i+1)}})-\frac{1}{2L^2\pi^2}(\gamma_{g\mathbf{i}}-\gamma_{g(\mathbf{i+1})})^2
\end{eqnarray}
in which
$1/L\sum_{\mathbf{k}}g_{\mathbf{ki}}^2=1-1/L\sum_{\mathbf{k}}h_{\mathbf{ki}}^2$
is used.


\end{document}